\documentclass{PoS}

\title{Neutron-Antineutron Operator Renormalization}

\ShortTitle{Neutron-Antineutron Operator Renormalization}

\author{Michael I. Buchoff\\
        University of Washinton, Institute for Nuclear Theory\\
        E-mail: \email{mbuchoff@u.washington.edu}}

\author{\speaker{Michael Wagman}\\
        University of Washington, Institute for Nuclear Theory\\
        E-mail: \email{mlwagman@uw.edu}}

		\abstract{Baryon number symmetry violating theories beyond the standard model with suppressed proton decay rates can be experimentally constrained by data on neutron-antineutron transition rates. In order to apply this constraints, theoretical predictions for the neutron-antineutron transition rates in various models must be available for comparison. Reliable predictions of transition rates between hadronic states must include non-perturbative quantum chromodynamic effects. These can be calculated in a model independent way by calculating six-quark operator matrix elements with lattice quantum chromodynamics. Preliminary lattice calculations have been performed, but operator renormalization effects must be included in order to match beyond the standard model calculations performed in $\bar{MS}$ renormalized perturbation theory with lattice regularized matrix element results. In particular, a perturbative calculation of the two-loop anomalous dimensions and one-loop renormalization scheme matching coefficients of these six-quark operators is necessary in order to determine leading order corrections at lattice matching scales. This describes our ongoing calculation of these perturbative operator renormalization effects.}

\FullConference{The 32nd International Symposium on Lattice Field Theory,\\
		23-28 June, 2014\\
		Columbia University New York, NY}

\usepackage{amsmath}
\newcommand{\avg}[1]{\left< #1 \right>} % for average
\let\bar=\smallbar % rename builtin \bar{} to \smallbar{}
\newcommand{\bar}[1]{\overline{#1}} % default to large bars

\begin{document}

\section{Introduction}

The apparent matter-antimatter asymmetry of the universe remains an outstanding mystery of physics that cannot be explained within the standard model (SM) \cite{Cohen:1993nk}. This disagreement between the SM and observation motivates the existence of beyond the standard model (BSM) theories with additional violation of $CP$ and baryon number ($B$) conservation. Many BSM models with additional symmetry violation can be constructed, and experimental guidance is necessary to constrain the space of viable BSM models and determine what theory describes our matter-dominated universe. 

Experimental measurements of, or bounds on, baryon number violation rates can only be directly compared to theoretical predictions of transition rates between hadronic states. Reliable calculations of hadronic observables must include the strong nuclear forces present, and performing top-down calculations of hadronic transition rates for every BSM model of interest is highly impractical. Instead, baryon number violating processes can be treated in a model-independent way by considering the standard model as an effective field theory that includes baryon number violating operators composed only of quarks and other SM degrees of freedom. The predictions of particular BSM models of interest can be straightforwardly parametrized in terms of local SM operators. Once the model-independent matrix elements of these operators have been determined, BSM models can be directly confronted with experimental constraints.

The most relevant baryon number violating operator that can be added to the standard model is a dimension 6 operator that permits proton decay. Experimental constraints on proton decay rates place stringent bounds on matrix elements of this operator and therefore the scale of $\Delta B = 1$ physics. Many phenomenologically attractive BSM theories include a suppression of these well-constrained $\Delta B = 1$ effects, see Ref. \cite{Mohapatra:2009wp} for a review. In these theories, the baryon number violating effects most accessible in low-energy experiments are typically $\Delta B = 2$ neutron-antineutron transitions. Neutron decay searches have a long history \cite{Takita:1986zm}, but extracting neutron transition rates from observations of nuclei remains challenging. Experiments using reactor neutron beams can provide clean measurements of the $n-\bar{n}$ transition rate \cite{BaldoCeolin:1994jz}, and current experimental bounds have the potential to be greatly improved with new measurements \cite{Kronfeld:2013uoa}. 

Turning precise experimental data on $n-\bar{n}$ transition rates into precise constraints on BSM physics presents theoretical challenges. Naive dimensional analysis alone cannot be used to reliably estimate transition rates--varying the value of $\Lambda_{QCD}$ used for naive dimensional analysis of a six-quark operator matrix element leads to a thousand-fold transition rate variation. The requisite nuclear matrix elements have also been estimated within the MIT bag model \cite{Rao:1982gt}, but there is no way to determine the theoretical uncertainty present.

The only technique currently available for turning experimental $n-\bar{n}$ transition rate data into constraints on BSM physics with quantified uncertainties is lattice quantum chromodynamics (LQCD). Preliminary LQCD calculations of the required matrix elements have been performed \cite{Buchoff:2012bm}. These simulations include large systemic uncertainties arising from the need to relate lattice regularized matrix elements defined at GeV scales to $\bar{MS}$ renormalized matrix elements defined at high scales where reliable BSM predictions have been made. This report describes an ongoing calculation of these operator renormalization effects.

\section{Operator Renormalization}

The Hamiltonian for $n-\bar{n}$ transitions can be expanded in a complete basis of six-quark operators, explicitly constructed below. Denoting these operators by $\mathcal{O}^{\mathbf{a}}$ the Wilson coefficients predicted by a particular BSM theory by $C^{\mathbf{a}}$, the Hamiltonian density for $n-\bar{n}$ transitions is expanded as
\begin{eqnarray}
  \mathcal{H}_{n\bar{n}} &=& C_0^{\mathbf{a}} \mathcal{O}_0^{\mathbf{a}} = C^{\mathbf{a}}(\mu) \mathcal{O}^{\mathbf{a}}(\mu),
\end{eqnarray}
where the first equality involves bare operators and coefficients and the second renormalized operators and coefficients defined at an arbitrary renormalization scale $\mu$. These are related by renormalization coefficients that are defined for a particular renormalization scheme and scale as
\begin{equation}
    \mathcal{O}_{scheme}^{\mathbf{a}}(\mu) = Z^{\mathbf{ab}}_{scheme}(\mu) \mathcal{O}_0^{\mathbf{b}}.
    \label{Zdef}
\end{equation}
BSM models can typically be used to perturbatively calculate Wilson coefficients of $\bar{MS}$ renormalized operators at high scales. Model independent relations between these coefficients and measurable $n-\bar{n}$ transition rates can only be provided once lattice regularized matrix elements extracted from LQCD simulations are related to $\bar{MS}$ renormalized matrix elements at high scales.

Since the $\bar{MS}$ renormalization prescription can only be applied to dimensionally regularized (dim reg) matrix elements, relating lattice and $\bar{MS}$ matrix elements necessitates the introduction of an intermediate renormalization scheme that can be applied to both lattice and dim reg matrix elements. The Regularization Invariant Momentum (RI-MOM) scheme is well suited for this \cite{Martinelli:1994ty}, and has been utilized for practical operator renormalization of kaon oscillation \cite{Aoki:2010pe} and proton decay \cite{Aoki:2013yxa} matrix elements. 

The RI-MOM renormalization scheme, explicitly defined below, is implemented by demanding that vertex functions calculated at a reference momentum $p_0$ are fixed to a particular value. This definition can be applied non-perturbatively to LQCD matrix elements in order to determine the coefficient matrix $Z_{RI}^{latt}(p_0)$ relating lattice regularized and RI-MOM renormalized operators. The RI-MOM scheme can also be applied perturbatively to dimensionally regularized matrix elements to determine $Z_{RI}^{cont}(p_0)$ and necessarily removes all poles in $D-4$ to produce finite renormalized matrix elements. This means that the ratio of $Z_{RI}^{cont}(p_0)$ to $Z_{\bar{MS}}(p_0)$ is a finite quantity that can be reliably calculated in perturbation theory provided $\alpha_s(p_0) \ll 1$.

Combining the matching factor given by this ratio with standard renormalization group analysis, the relation between $\bar{MS}$ operators at arbitrary scales and lattice operators with quark momenta $p_0$ is given by
\begin{eqnarray}
  \nonumber
	\mathcal{O}_{\bar{MS}}(\mu) &=& \left( \frac{Z_{\bar{MS}}(p_0)}{Z_{RI}^{cont}(p_0)} \right)Z_{RI}^{latt}(p_0)\left[ \frac{\alpha_s(\mu)}{\alpha_s(p_0)} \right]^{-\gamma_0/2\beta_0}\left[ 1 + \left( \frac{\beta_1 \gamma_0}{2\beta_0^2} - \frac{\gamma_1}{2\beta_0} \right)\frac{\alpha_s(\mu) - \alpha_s(p_0)}{4\pi} \right]\mathcal{O}_{latt},\\
        \label{master}
\end{eqnarray}
where $\beta_0$ and $\beta_1$ are the one-loop and two-loop coefficients of the QCD $\beta$-function and $\gamma_0$ and $\gamma_1$ are the one-loop and two-loop coefficients of the $\Delta B = 2$ operator anomalous dimension matrix. 

Keeping both discretization and perturbative errors parametrically suppressed requires the hierarchy of scales $\Lambda_{QCD}\ll p_0 \ll a^{-1}$ where $a$ is the LQCD lattice spacing. BSM matching calculations are performed at much higher scales than present inverse lattice spacings, and so Eq. \ref{master} will be generally applied to cases where $\alpha_s(\mu) \ll \alpha_s(p_0)$. In this case both one-loop contributions to the matching factor $\left( \frac{Z^{\bar{MS}}(p_0)}{Z^{MOM}_{cont}(p_0)} \right)$ and the two-loop anomalous dimension $\gamma_1$ provide $O(\alpha_s(p_0))$ operator renormalization effects. The one-loop anomalous dimension has been known for some time \cite{Caswell:1982qs}, but both the one-loop matching factor and two-loop anomalous dimensions are unknown and form the targets of our investigation.

\section{Operators and Vertex Functions}\label{ops}

Six-quark operators that are Lorentz invariant, local, standard model gauge invariant, and have the correct flavor structure to describe $n-\bar{n}$ transitions can be expressed in a basis of operators of the form\footnote{We thank Brian Tiburzi and Sergey Syritsyn for helpful discussions on the symmetry transformation properties of six-quark operators and optimal operator basis construction.}
\begin{eqnarray}
    \label{opdef}
    \mathcal{O}^1_{\chi_1\chi_2\chi_3} &=& (u^T_i C P_{\chi_1} u_j)(d^T_j C P_{\chi_2} d_l)(d^T_m C P_{\chi_3} d_n)T_{\{ij\}\{kl\}\{mn\}},\\\nonumber
    \mathcal{O}^2_{\chi_1\chi_2\chi_3} &=& (u^T_i C P_{\chi_1} d_j)(u^T_k C P_{\chi_2} d_l)(d^T_m C P_{\chi_3} d_n)T_{\{ij\}\{kl\}\{mn\}},\\\nonumber
    \mathcal{O}^3_{\chi_1\chi_2\chi_3} &=& (u^T_i C P_{\chi_1} d_j)(u^T_k C P_{\chi_2} d_l)(d^T_m C P_{\chi_3} d_n)T_{[ij][kl]\{mn\}},
\end{eqnarray}
where $\chi_1,\chi_2,\dots = \pm$ are chirality labels with $P_{\pm} = \frac{1}{2}(1\pm \gamma_5)$, $C$ is the charge conjugation matrix, and $i,j,k\cdots$ above are color labels. The only tensors that can combine six quarks of this flavor structure into color singlets are
\begin{eqnarray}
	T_{\{ij\}\{kl\}\{mn\}} &=& \varepsilon_{ikm}\varepsilon_{jln} + \varepsilon_{jkm}\varepsilon_{iln} + \varepsilon_{ilm}\varepsilon_{jkn}  + \varepsilon_{ikn}\varepsilon_{jlm},\\\nonumber
	T_{[ij][kl]\{mn\}} &=& \varepsilon_{ijm}\varepsilon_{kln} + \varepsilon_{ijn}\varepsilon_{klm},
\end{eqnarray}
where $\varepsilon_{ijk}$ is the Levi-Civita symbol. These tensors are symmetric $\{ij\}$ or antisymmetric $[ij]$ in each pair of indices as shown. Operators with more complicated Dirac matrix structures are not independent and can be expressed as linear combinations of those above through Fierz transformations. The trivial constraints $\mathcal{O}^1_{\chi LR} = \mathcal{O}^1_{\chi RL}$, $\mathcal{O}^{2,3}_{RL\chi}=\mathcal{O}^{2,3}_{LR\chi}$ and non-trivial constraints $\mathcal{O}^2_{\chi\chi\chi^\prime} - \mathcal{O}^1_{\chi\chi\chi^\prime} = 3\mathcal{O}^3_{\chi\chi\chi^\prime}$ immediately reduce the number of independent operators of this form from 24 to 14. Parity further reduces the number of independent operators to 7. Chiral symmetry provides further constraints, but maintaining a complete operator basis when utilizing chiral symmetry breaking lattice regularizations requires a basis of 7 independent operators such as
\begin{eqnarray}
    \mathcal{O}^\mathbf{a} &=& \{\mathcal{O}^1_{RLL},\; \mathcal{O}^2_{LLR},\; \mathcal{O}^2_{LRL},\; \mathcal{O}^2_{RRR},\; \mathcal{O}^3_{LLR},\; \mathcal{O}^3_{LRR},\; \mathcal{O}^3_{RRR}\}.
\end{eqnarray}

Vertex functions for each operator can be constructed (in both perturbation theory and LQCD) by Wick contracting with neutron and antineutron interpolating operators
\begin{eqnarray}
    (\Lambda^{\bf{a}})^{\alpha\beta\gamma\delta\eta\sigma}_{ijklmn} = \left.\avg{\bar{u}^\alpha_i \bar{d}^\gamma_k \bar{d}^\delta_l \left({}^{\bf{a}}\mathcal{O}\right) \bar{u}^\beta_j \bar{d}^\eta_m \bar{d}^\sigma_n }\right|_{amp},
\end{eqnarray}
where the subscript amp refers to the prescription of amputating external legs by taking
\begin{eqnarray}
	\bar{q}_i^\alpha \rightarrow \bar{q}_{i^\prime}^{\alpha^\prime} \avg{(S_q^{-1})_{i^\prime i}^{\alpha^\prime\alpha}},
\end{eqnarray}
where $S_q$ is the propagator for quark flavor $q$. The tree-level vertex functions for these operators are given explicitly by
\begin{eqnarray}
    &&\left(\Lambda^{1}_{(0)}\right)_{ijklmn}^{\alpha\beta\gamma\delta\eta\sigma} = \avg{\bar{d}_n^\sigma \bar{d}_m^\eta  \bar{d}_l^\delta \bar{d}_k^\gamma  \bar{u}_j^\beta \bar{u}_i^\alpha u_{i^\prime}^{\alpha^\prime}u_{j^\prime}^{\beta^\prime} d_{k^\prime}^{\gamma^\prime} d_{l^\prime}^{\delta^\prime} d_{m^\prime}^{\eta^\prime}d_{n^\prime}^{\sigma^\prime} }X_1^{\alpha^\prime\beta^\prime}X_2^{\gamma^\prime\delta^\prime}X_3^{\eta^\prime\sigma^\prime}T_{\{i^\prime j^\prime\}\{k^\prime l^\prime\}\{m^\prime n^\prime\}}\\\nonumber
    &&\hspace{10pt}=8T_{\{ij\}\{kl\}\{mn\}}\big[X_1^{\alpha \beta}X_2^{\gamma \delta}  X_3^{\eta \sigma} + X_1^{\alpha \beta}X_2^{\eta \sigma}  X_3^{\gamma \delta}  \big]+8T_{\{ij\}\{km\}\{ln\}}\big[X_1^{\alpha \beta}X_2^{\eta \gamma}  X_3^{\delta \sigma} + X_1^{\alpha \beta}X_2^{\delta \sigma}  X_3^{\eta \gamma} \big]\nonumber\\
    &&\hspace{20pt}+8T_{\{ij\}\{kn\}\{lm\}}\big[X_1^{\alpha \beta}X_2^{\gamma \sigma}  X_3^{\delta \eta} + X_1^{\alpha \beta}X_2^{\delta \eta}  X_3^{\gamma \sigma}\big], \nonumber\\
    &&\left(\Lambda^{2,3}_{(0)}\right)_{ijklmn}^{\alpha\beta\gamma\delta\eta\sigma} = -\avg{\bar{d}_n^\sigma \bar{d}_m^\eta  \bar{d}_l^\delta \bar{d}_k^\gamma  \bar{u}_j^\beta \bar{u}_i^\alpha u_{i^\prime}^{\alpha^\prime}u_{j^\prime}^{\beta^\prime} d_{k^\prime}^{\gamma^\prime} d_{l^\prime}^{\delta^\prime} d_{m^\prime}^{\eta^\prime}d_{n^\prime}^{\sigma^\prime} }X_1^{\alpha^\prime\gamma^\prime}X_2^{\beta^\prime\delta^\prime}X_3^{\eta^\prime\sigma^\prime}T_{(i^\prime k^\prime)(j^\prime l^\prime)\{m^\prime n^\prime\}}\\\nonumber
    &&\hspace{10pt}= 2T_{(ik)(jl)\{mn\}}\big[X_1^{\alpha \gamma}X_2^{\beta \delta}  X_3^{\sigma \eta} + X_1^{\beta \delta}X_2^{\alpha \gamma}  X_3^{\sigma \eta}  \big]+2T_{(ik)(jm)\{ln\}}\big[X_1^{\alpha \gamma}X_2^{\beta \eta}  X_3^{\delta \sigma} + X_1^{\beta \eta}X_2^{\alpha \gamma}  X_3^{\delta \sigma} \big]\nonumber\\
    &&\hspace{20pt} +2T_{(ik)(jn)\{lm\}}\big[X_1^{\alpha \gamma}X_2^{\beta \sigma}  X_3^{\eta \delta} + X_1^{\beta \sigma}X_2^{\alpha \gamma}  X_3^{\eta \delta}  \big]+2T_{(il)(jk)\{mn\}}\big[X_1^{\alpha \delta}X_2^{\beta \gamma}  X_3^{\eta \sigma} + X_1^{\beta \gamma}X_2^{\alpha \delta}  X_3^{\eta \sigma} \big]\nonumber\\
    &&\hspace{20pt} +2T_{(il)(jm)\{kn\}}\big[X_1^{\alpha \delta}X_2^{\beta \eta}  X_3^{\sigma \gamma} + X_1^{\beta \eta}X_2^{\alpha \delta}  X_3^{\sigma \gamma}  \big]+2T_{(il)(jn)\{km\}}\big[X_1^{\alpha \delta}X_2^{\beta \sigma}  X_3^{\gamma \eta} + X_1^{\beta \sigma}X_2^{\alpha \delta}  X_3^{\gamma \eta} \big]\nonumber\\
    &&\hspace{20pt}+2T_{(im)(jk)\{ln\}}\big[X_1^{\alpha \eta}X_2^{\beta \gamma}  X_3^{\sigma \delta} + X_1^{\beta \gamma}X_2^{\alpha \eta}  X_3^{\sigma \delta}  \big]+2T_{(im)(jl)\{kn\}}\big[X_1^{\alpha \eta}X_2^{\beta \delta}  X_3^{\gamma \sigma} + X_1^{\beta \delta}X_2^{\alpha \eta}  X_3^{\gamma \sigma} \big]\nonumber\\
    &&\hspace{20pt}+2T_{(im)(jn)\{kl\}}\big[X_1^{\alpha \eta}X_2^{\beta \sigma}  X_3^{\delta \gamma} + X_1^{\beta \sigma}X_2^{\alpha \eta}  X_3^{\delta \gamma}  \big]+2T_{(in)(jk)\{lm\}}\big[X_1^{\alpha \sigma}X_2^{\beta \gamma}  X_3^{\delta \eta} + X_1^{\beta \gamma}X_2^{\alpha \sigma}  X_3^{\delta \eta} \big]\nonumber\\\nonumber
    &&\hspace{20pt}+2T_{(in)(jl)\{km\}}\big[X_1^{\alpha \sigma}X_2^{\beta \delta}  X_3^{\eta \gamma} + X_1^{\beta \delta}X_2^{\alpha \sigma}  X_3^{\eta \gamma}  \big]+2T_{(in)(jm)\{kl\}}\big[X_1^{\alpha \sigma}X_2^{\beta \eta}  X_3^{\gamma \delta} + X_1^{\beta \eta}X_2^{\alpha \sigma}  X_3^{\gamma \delta} \big],
\end{eqnarray}
where $X_i = CP_{\chi_i}$, $(ij)$ denotes $\{ij\}$ for ${}^2\Lambda$ and $[ij]$ for ${}^3\Lambda$, and we have suppressed chiral indices on ${}^{\mathbf{a}}\Lambda$.

The RI-MOM renormalization condition is formally defined by a condition on these vertex functions evaluated with all quarks at a reference (Euclidean) momentum $p_0$ and $\mu^2 = p_0^2$
\begin{eqnarray}
    \label{MOM}
    \left.Z_{RI}^{\bf{ab}}\left( Z_{RI}^q \right)^{-3} \left(P^{\bf{b}}\right)^{\alpha\beta\gamma\delta\eta\sigma}_{ijklmn} \left(\Lambda^{\bf{c}}\right)^{\alpha\beta\gamma\delta\eta\sigma}_{ijklmn}\right|_{\mu^2 = p_0^2} = \delta^{\bf{ac}},
\end{eqnarray}
where $Z_{MOM}^q$ is the RI-MOM wavefunction renormalization factor defined in Ref \cite{Martinelli:1994ty}, $Z_{MOM}^{\mathbf{ab}}$ is the matrix of renormalization coefficients appearing in Eqs. \eqref{Zdef}, \eqref{master}, and $P^{\mathbf{a}}$ projects onto the tree-level vertex function $\Lambda_{(0)}^{\mathbf{a}}$ according to the defining relation
\begin{eqnarray}
    \label{projectordef}
    (P^{\bf{a}})_{ijklmn}^{\alpha\beta\gamma\delta\eta\sigma}\left(\Lambda_{(0)}^{\bf{b}}\right)_{ijklmn}^{\alpha\beta\gamma\delta\eta\sigma} = \delta^{\bf{ab}}.
\end{eqnarray}
There is some freedom in constructing projectors that satisfy Eq. \eqref{projectordef}. The set we have constructed is given explicitly by
\begin{eqnarray}
    \label{projectors}
    \left(P_{RLL}^1\right)^{\alpha \beta\gamma\delta\sigma}_{ijklmn}&=&\frac{1}{55296}T_{\{ij\}\{kl\}\{mn\}}X_R^{\alpha \beta}X_L^{\gamma \delta}  X_L^{\eta \sigma}\\\nonumber
    \left(P_{LRL}^2\right)^{\alpha \beta\gamma\delta\sigma}_{ijklmn}&=&\frac{1}{13824}\big[T_{\{ij\}\{kl\}\{mn\}}X_R^{\alpha \delta}X_L^{\gamma \beta}  X_L^{\eta \sigma}+6T_{[ij][kl]\{mn\}}X_R^{\alpha \delta}X_L^{\gamma \sigma}  X_L^{\eta \beta}\big]\\\nonumber
    \left(P_{RLL}^3\right)^{\alpha \beta\gamma\delta\sigma}_{ijklmn}&=&\frac{1}{4608}\big[-T_{\{ij\}\{kl\}\{mn\}}X_R^{\alpha \delta}X_L^{\gamma \beta}  X_L^{\eta \sigma}+2T_{[ij][kl]\{mn\}}X_R^{\alpha \delta}X_L^{\gamma \sigma}  X_L^{\eta \beta}\big]\\\nonumber
    \left(P_{LLL}^2\right)^{\alpha \beta\gamma\delta\sigma}_{ijklmn}&=&\frac{1}{55296}\big[T_{\{ij\}\{kl\}\{mn\}}X_L^{\alpha \beta}X_L^{\gamma \delta}  X_L^{\eta \sigma}+3T_{[ij][kl]\{mn\}}X_L^{\alpha \delta}X_L^{\gamma \beta}  X_L^{\eta \sigma}\big]\\\nonumber
    \left(P_{LLL}^3\right)^{\alpha \beta\gamma\delta\sigma}_{ijklmn}&=&\frac{1}{18432}\big[-T_{\{ij\}\{kl\}\{mn\}}X_L^{\alpha \beta}X_L^{\gamma \delta}  X_L^{\eta \sigma}+T_{[ij][kl]\{mn\}}X_L^{\alpha \delta}X_L^{\gamma \beta}  X_L^{\eta \sigma}\big]\\\nonumber
    \left(P_{RRL}^2\right)^{\alpha \beta\gamma\delta\sigma}_{ijklmn}&=&\frac{1}{18432}\big[T_{\{ij\}\{kl\}\{mn\}}X_R^{\alpha \beta}X_R^{\gamma \delta}  X_L^{\eta \sigma}+2T_{[ij][kl]\{mn\}}X_R^{\alpha \delta}X_R^{\gamma \beta}  X_L^{\eta \sigma}\big]\\\nonumber
    \left(P_{RRL}^3\right)^{\alpha \beta\gamma\delta\sigma}_{ijklmn}&=&\frac{1}{18432}\big[-3T_{\{ij\}\{kl\}\{mn\}}X_R^{\alpha \beta}X_R^{\gamma \delta}  X_L^{\eta \sigma}+2T_{[ij][kl]\{mn\}}X_R^{\alpha \delta}X_R^{\gamma \beta}  X_L^{\eta \sigma}\big].
\end{eqnarray}  

Projectors satisfying Eq. \eqref{projectordef} are useful for decomposing matrix elements into a desired operator basis in more general contexts than enforcing the RI-MOM condition Eq. \eqref{MOM}. In particular, we can express the $\bar{MS}$ renormalization condition as
\begin{eqnarray}
    \label{MSbar}
	Z_{\bar{MS}}^{\mathbf{ab}}\left( Z_{\bar{MS}}^q \right)^{-3} \left(P^{\mathbf{b}}\right)^{\alpha\beta\gamma\delta\eta\sigma}_{ijklmn} \left(\Lambda_{pole}^{\bf{c}}\right)^{\alpha\beta\gamma\delta\eta\sigma}_{ijklmn} = \delta^{\mathbf{ac}},
\end{eqnarray}
where $\Lambda_{pole}^{\mathbf{a}}$ denotes the dim reg vertex function contributions proportional to $1/\bar{\varepsilon} = 2/(4-D) - \gamma_E + \ln 4\pi$, $Z_{\bar{MS}}$ is the usual $\bar{MS}$ quark wavefunction renormalization factor, and we have suppressed dependence on the renormalization scale $\mu$.

Using projectors to decompose matrix elements into a particular operator basis is essential for LQCD relations where the analytic relations between full and tree-level vertex functions are unknown. Given the complicated tensor structures of our vertex functions, the projector definitions of $Z_{RI}$ and $Z_{\bar{MS}}$ have practical advantages for our perturbative calculations as well. Most importantly, applying the projector definitions greatly simplifies the tensor decomposition of the many spin, color, and flavor structures produced by one-loop and especially two-loop diagrams in dim reg.

\section{Perturbative Calculation}

A calculation of the one-loop RI-MOM/$\bar{MS}$ matching factor includes 15 one-loop Feynman diagrams. These can be straightforwardly expressed as contractions of four independent tensor integrals, two for Feynman gauge and two more for general $R_\xi$ gauge needed for consistency with LQCD simulations performed in Landau gauge. 

The contractions of these tensor integrals with the Dirac structures of the diagrams introduces subtleties, however. Applying the projectors to structures such as $\frac{1}{\bar{\varepsilon}}\left[ \sigma_{\mu\nu}P_{\chi_1} \right]^{\alpha\beta}\left[ \sigma_{\mu\nu}P_{\chi_2} \right]^{\gamma\delta}P_{\chi_3}^{\eta\sigma}$ introduces traces of 4 $\gamma^\mu$ times $\gamma_5$ that are ill-defined in naive dimensional regularization. While these structures can be related to our original operator basis through Fierz transformations, these Fierz transforms are only valid in four dimensions. To consistently account for the fact that our operator basis is only complete in $D=4$ we follow the approach of Ref. \cite{Buras:1989xd} and introduce evanescent operators proportional to $D-4$ that complete our basis in $D$ dimensions. Both the one-loop matching factors and two-loop anomalous dimensions depend on the basis of evanescent operators chosen and are ambiguous until this basis is specified.

320 two-loop diagrams contribute to the $\bar{MS}$ anomalous dimension matrix. Standard multiloop techniques can be used to perform tensor reductions of the independent tensor integrals appearing and to use integration by parts algorithms to recursively decompose the resulting scalar integrals into a minimal set that can be evaluated straightforwardly \cite{Passarino:1978jh, Chetyrkin:1981qh, Tkachov:1984xk}. Slightly more than half of the independent two-loop topologies can be treated as simple generalizations of the diagram topologies appearing in four-quark operator renormalization calculations, but the remaining diagrams involving gluon exchange between all three spin-singlet diquark structures in our operators have no direct four-quark analogs. Proper treatment of evanescent operators in these diagrams in particular requires the introduction of new evanescent operator structures that do not appear in four-quark operator calculations.

\section{Conclusion}

Extracting predictions of $n-\bar{n}$ transition rates with quantified uncertainties from BSM theories relevant for baryogensis requires LQCD simulation as well as perturbative operator renormalization analysis. Accurately calculating relations between lattice regularized and $\bar{MS}$ renormalized matrix elements requires both non-perturbative renormalization of LQCD matrix elements in a regularization invariant scheme such as RI-MOM and perturbative matching between RI-MOM and $\bar{MS}$ renormalization factors. In addition to this one-loop renormalization scheme matching calculation, the two-loop $\bar{MS}$ anomalous dimension matrix is needed for accurate relations between BSM matching calculations and LQCD simulations at different scales. 

We have begun an ongoing calculation of both of these perturbative quantities using suitably generalized four-quark operator renormalization techniques as well as operator projectors that simplify two-loop tensor decompositions. Completing this calculation will remove a significant systematic uncertainty in using $n-\bar{n}$ transition rate measurements to constrain BSM physics. Our two-loop calculation is well underway, and we will publish results on both the one-loop matching factors and two-loop anomalous dimension matrix once the two-loop calculation is complete.


\begin{thebibliography}{99}
	%\cite{Cohen:1993nk}
  \bibitem{Cohen:1993nk} 
	  A.~G.~Cohen, D.~B.~Kaplan and A.~E.~Nelson,
	    %``Progress in electroweak baryogenesis,''
	    Ann.\ Rev.\ Nucl.\ Part.\ Sci.\  {\bf 43}, 27 (1993) [hep-ph/9302210].
		    %%CITATION = HEP-PH/9302210;%%
		    %658 citations counted in INSPIRE as of 13 Dec 2014

%\cite{Mohapatra:2009wp}
		\bibitem{Mohapatra:2009wp} 
		    R.~N.~Mohapatra,
			  %``Neutron-Anti-Neutron Oscillation: Theory and Phenomenology,''
			  J.\ Phys.\ G {\bf 36}, 104006 (2009)
			    [arXiv:0902.0834 [hep-ph]].
				  %%CITATION = ARXIV:0902.0834;%%
				  %37 citations counted in INSPIRE as of 13 Dec 2014

				%\cite{Takita:1986zm}
				\bibitem{Takita:1986zm} 
				  M.~Takita {\it et al.}  [KAMIOKANDE Collaboration],
				    %``A Search for Neutron - Anti-neutron Oscillation in a $^{16}$O Nucleus,''
				    Phys.\ Rev.\ D {\bf 34}, 902 (1986).
					  %%CITATION = PHRVA,D34,902;%%
					  %64 citations counted in INSPIRE as of 26 Jan 2015

%\cite{BaldoCeolin:1994jz}
			  \bibitem{BaldoCeolin:1994jz} 
				  M.~Baldo-Ceolin, P.~Benetti, T.~Bitter, F.~Bobisut, E.~Calligarich, R.~Dolfini, D.~Dubbers and P.~El-Muzeini {\it et al.},
				    %``A New experimental limit on neutron - anti-neutron oscillations,''
				    Z.\ Phys.\ C {\bf 63}, 409 (1994).
					  %%CITATION = ZEPYA,C63,409;%%
					  %91 citations counted in INSPIRE as of 13 Dec 2014

%\cite{Mohapatra:1980qe}
											%\cite{Kronfeld:2013uoa}
										  \bibitem{Kronfeld:2013uoa} 
											  A.~S.~Kronfeld, R.~S.~Tschirhart, U.~Al-Binni, W.~Altmannshofer, C.~Ankenbrandt, K.~Babu, S.~Banerjee and M.~Bass {\it et al.},
											    %``Project X: Physics Opportunities,''
											    arXiv:1306.5009 [hep-ex].
												  %%CITATION = ARXIV:1306.5009;%%
												  %43 citations counted in INSPIRE as of 13 Dec 2014

												%\cite{Rao:1982gt}
											  \bibitem{Rao:1982gt} 
												  S.~Rao and R.~Shrock,
												    %``$n \leftrightarrow \bar{n}$ Transition Operators and Their Matrix Elements in the {MIT} Bag Model,''
												    Phys.\ Lett.\ B {\bf 116}, 238 (1982).
													  %%CITATION = PHLTA,B116,238;%%
													  %54 citations counted in INSPIRE as of 26 Jan 2015




													%\cite{Buchoff:2012bm}
												  \bibitem{Buchoff:2012bm} 
													  M.~I.~Buchoff, C.~Schroeder and J.~Wasem,
													    %``Neutron-antineutron oscillations on the lattice,''
													    PoS LATTICE {\bf 2012}, 128 (2012)
														  [arXiv:1207.3832 [hep-lat]].
														    %%CITATION = ARXIV:1207.3832;%%
														    %13 citations counted in INSPIRE as of 13 Dec 2014

														  %\cite{Martinelli:1994ty}
														\bibitem{Martinelli:1994ty} 
														    G.~Martinelli, C.~Pittori, C.~T.~Sachrajda, M.~Testa and A.~Vladikas,
															  %``A General method for nonperturbative renormalization of lattice operators,''
															  Nucl.\ Phys.\ B {\bf 445}, 81 (1995)
															    [hep-lat/9411010].
																  %%CITATION = HEP-LAT/9411010;%%
																  %563 citations counted in INSPIRE as of 13 Dec 2014

																%\cite{Aoki:2010pe}
															  \bibitem{Aoki:2010pe} 
																  Y.~Aoki, R.~Arthur, T.~Blum, P.~A.~Boyle, D.~Brommel, N.~H.~Christ, C.~Dawson and T.~Izubuchi {\it et al.},
																    %``Continuum Limit of $B_K$ from 2+1 Flavor Domain Wall QCD,''
																    Phys.\ Rev.\ D {\bf 84}, 014503 (2011)
																	  [arXiv:1012.4178 [hep-lat]].
																	    %%CITATION = ARXIV:1012.4178;%%
																	    %51 citations counted in INSPIRE as of 13 Dec 2014

																%\cite{Aoki:2013yxa}
															  \bibitem{Aoki:2013yxa} 
																  Y.~Aoki, E.~Shintani and A.~Soni,
																    %``Proton decay matrix elements on the lattice,''
																    Phys.\ Rev.\ D {\bf 89}, 014505 (2014)
																	  [arXiv:1304.7424 [hep-lat]].
																	    %%CITATION = ARXIV:1304.7424;%%
																	    %14 citations counted in INSPIRE as of 13 Dec 2014

																	  %\cite{Caswell:1982qs}
																	\bibitem{Caswell:1982qs} 
																	    W.~E.~Caswell, J.~Milutinovic and G.~Senjanovic,
																		  %``Matter - Antimatter Transition Operators: A Manual For Modeling,''
																		  Phys.\ Lett.\ B {\bf 122}, 373 (1983).
																		    %%CITATION = PHLTA,B122,373;%%
																		    %7 citations counted in INSPIRE as of 13 Dec 2014

                                                                                                                                                  %\cite{Buras:1989xd}
                                                                                                                                              \bibitem{Buras:1989xd} 
                                                                                                                                                    A.~J.~Buras and P.~H.~Weisz,
                                                                                                                                                      %``QCD Nonleading Corrections to Weak Decays in Dimensional Regularization and 't Hooft-Veltman Schemes,''
                                                                                                                                                      Nucl.\ Phys.\ B {\bf 333}, 66 (1990).
                                                                                                                                                        %%CITATION = NUPHA,B333,66;%%
                                                                                                                                                        %267 citations counted in INSPIRE as of 14 Dec 2014 

                                                                                                                                                                %\cite{Passarino:1978jh}
                                                                                                                                                            \bibitem{Passarino:1978jh} 
                                                                                                                                                                  G.~Passarino and M.~J.~G.~Veltman,
                                                                                                                                                                    %``One Loop Corrections for e+ e- Annihilation Into mu+ mu- in the Weinberg Model,''
                                                                                                                                                                    Nucl.\ Phys.\ B {\bf 160}, 151 (1979).
                                                                                                                                                                      %%CITATION = NUPHA,B160,151;%%
                                                                                                                                                                      %1801 citations counted in INSPIRE as of 14 Dec 2014

                                                                                                                                                      %\cite{Chetyrkin:1981qh}
                                                                                                                                                  \bibitem{Chetyrkin:1981qh} 
                                                                                                                                                        K.~G.~Chetyrkin and F.~V.~Tkachov,
                                                                                                                                                          %``Integration by Parts: The Algorithm to Calculate beta Functions in 4 Loops,''
                                                                                                                                                          Nucl.\ Phys.\ B {\bf 192}, 159 (1981).
                                                                                                                                                            %%CITATION = NUPHA,B192,159;%%
                                                                                                                                                            %958 citations counted in INSPIRE as of 14 Dec 2014

                                                                                                                                                          %\cite{Tkachov:1984xk}
                                                                                                                                                      \bibitem{Tkachov:1984xk} 
                                                                                                                                                            F.~V.~Tkachov,
                                                                                                                                                              %``An Algorithm For Calculating Multiloop Integrals,''
                                                                                                                                                              Theor.\ Math.\ Phys.\  {\bf 56}, 866 (1983)
                                                                                                                                                                [Teor.\ Mat.\ Fiz.\  {\bf 56}, 350 (1983)].
                                                                                                                                                                  %%CITATION = TMPHA,56,866;%%
                                                                                                                                                                  %40 citations counted in INSPIRE as of 14 Dec 2014


\end{thebibliography}
\end{document}